\documentclass[aps,showpacs,showkeys,amsmath,amssymb,pra,reprint,superscriptaddress,lengthcheck]{revtex4-1}

\usepackage{graphicx}
\usepackage{dcolumn}
\usepackage{bm}
\usepackage{color}
\usepackage{amsmath}
\usepackage{amssymb}
\usepackage{gensymb}
\usepackage{slashed}
\usepackage{soul}
\usepackage{hyperref}
\hypersetup{colorlinks=true,
            linkcolor=blue,
            urlcolor=blue,
            citecolor=blue}

\definecolor{aogreen}{rgb}{0.0, 0.5, 0.0}
\definecolor{RiaanGreen}{RGB}{45,119,0}

\definecolor{revisedcolor}{RGB}{0,100,20}

\definecolor{Bcolor}{RGB}{10,200,10}
\definecolor{Jcolor}{RGB}{20,20,200}
\definecolor{Ccolor}{RGB}{200,20,20}
\definecolor{Qcolor}{RGB}{50,200,200}

\begin{document}
@book{Budker:2013,
  title={Optical Magnetometry},
  author={edited by D. Budker and D. F. Jackson Kimball},
  year={2013},
  publisher={Cambridge University Press, Cambridge, England}
}

@article{Budker/N:2007,
  title={Optical magnetometry},
  author={Budker, Dmitry and Romalis, Michael},
  journal={Nat. Phys.},
  volume={3},
  pages={227--234},
  year={2007},
  doi={10.1038/nphys566}
}

@article{Johnson/PMB:2013,
doi = {10.1088/0031-9155/58/17/6065},
url = {https://dx.doi.org/10.1088/0031-9155/58/17/6065},
year = {2013},
month = {aug},
publisher = {IOP Publishing},
volume = {58},
number = {17},
pages = {6065},
author = {Cort N Johnson and P D D Schwindt and M Weisend},
title = {Multi-sensor magnetoencephalography with atomic magnetometers},
journal = { Phys. Med. Biol.},
}

@article{Vasilakis/PRL:2009,
  title = {Limits on New Long Range Nuclear Spin-Dependent Forces Set with a $\mathbf{K}\mathrm{\text{\ensuremath{-}}}^{3}\mathrm{He}$ Comagnetometer},
  author = {Vasilakis, G. and Brown, J. M. and Kornack, T. W. and Romalis, M. V.},
  journal = {Phys. Rev. Lett.},
  volume = {103},
  issue = {26},
  pages = {261801},
  numpages = {4},
  year = {2009},
  month = {Dec},
  publisher = {American Physical Society},
  doi = {10.1103/PhysRevLett.103.261801},
  url = {https://link.aps.org/doi/10.1103/PhysRevLett.103.261801}
}

@article{Dang/APL:2010,
    author = {Dang, H. B. and Maloof, A. C. and Romalis, M. V.},
    title = "{Ultrahigh sensitivity magnetic field and magnetization measurements with an atomic magnetometer}",
    journal = { Appl. Phys. Lett.},
    volume = {97},
    number = {15},
    pages = {151110},
    year = {2010},
    month = {10},
    issn = {0003-6951},
    doi = {10.1063/1.3491215},
    url = {https://doi.org/10.1063/1.3491215}
}

@article{Shah/NP:2007,
  title={Subpicotesla atomic magnetometry with a microfabricated vapour cell},
  author={Shah, V. and Knappe, S. and Schwindt, P. D. D. and Kitching, J.},
  journal={Nat. Photonics},
  volume={1},
  pages={649},
  year={2007},
  doi={10.1038/nphoton.2007.201}
}

@article{Rubinsztein-Dunlop/JO:2017,
doi = {10.1088/2040-8978/19/1/013001},
url = {https://dx.doi.org/10.1088/2040-8978/19/1/013001},
year = {2017},
month = {nov},
publisher = {IOP Publishing},
volume = {19},
number = {1},
pages = {013001},
author = {Halina Rubinsztein-Dunlop and Andrew Forbes and M V Berry and M R Dennis and David L Andrews and Masud Mansuripur and Cornelia Denz and Christina Alpmann and Peter Banzer and Thomas Bauer and Ebrahim Karimi and Lorenzo Marrucci and Miles Padgett and Monika Ritsch-Marte and Natalia M Litchinitser and Nicholas P Bigelow and C Rosales-Guzmán and A Belmonte and J P Torres and Tyler W Neely and Mark Baker and Reuven Gordon and Alexander B Stilgoe and Jacquiline Romero and Andrew G White and Robert Fickler and Alan E Willner and Guodong Xie and Benjamin McMorran and Andrew M Weiner},
title = {Roadmap on structured light},
journal = {J. Opt.},
}

@book{Gbur:2017,
  title={Singular optics},
  author={Gbur, Gregory J},
  year={2017},
  publisher={CRC press}
}

@article{Castellucci/PRL:2021,
  title = {Atomic Compass: Detecting 3D Magnetic Field Alignment with Vector Vortex Light},
  author = {Castellucci, Francesco and Clark, Thomas W. and Selyem, Adam and Wang, Jinwen and Franke-Arnold, Sonja},
  journal = {Phys. Rev. Lett.},
  volume = {127},
  issue = {23},
  pages = {233202},
  numpages = {6},
  year = {2021},
  month = {Nov},
  publisher = {American Physical Society},
  doi = {10.1103/PhysRevLett.127.233202},
  url = {https://link.aps.org/doi/10.1103/PhysRevLett.127.233202}
}

@article{Qiu/PR:2021,
author = {Shuwei Qiu and Jinwen Wang and Francesco Castellucci and Mingtao Cao and Shougang Zhang and Thomas W. Clark and Sonja Franke-Arnold and Hong Gao and Fuli Li},
journal = {Photon. Res.},
keywords = {CCD cameras; Cylindrical vector beams; Nitrogen vacancy centers; Optical fields; Tunable diode lasers; Vector beams},
number = {12},
pages = {2325--2331},
publisher = {Optica Publishing Group},
title = {Visualization of magnetic fields with cylindrical vector beams in a warm atomic vapor},
volume = {9},
month = {Dec},
year = {2021},
url = {https://opg.optica.org/prj/abstract.cfm?URI=prj-9-12-2325},
doi = {10.1364/PRJ.418522},
}

@article{Sun/OE:23,
author = {Yujie Sun and Zhaoying Wang},
journal = {Opt. Express},
keywords = {Distributed feedback lasers; Light transmission; Optical fields; Spatial light modulators; Structured light; Vector beams},
number = {10},
pages = {15409--15422},
publisher = {Optica Publishing Group},
title = {Optically polarized selective transmission of a fractional vector vortex beam by the polarized atoms with external magnetic fields},
volume = {31},
month = {May},
year = {2023},
url = {https://opg.optica.org/oe/abstract.cfm?URI=oe-31-10-15409},
doi = {10.1364/OE.487426},
abstract = {},
}

@article{Cai/LPR:24,
author = {Cai, Guoan and Tian, Ke and Wang, Zhaoying},
title = {Thermal Atomic Compass Based on Radially Polarized Beam},
journal = {Laser \& Photonics Reviews},
volume = {18},
number = {11},
pages = {2400465},
keywords = {polarization selection absorption, radially polarized beam, thermal atomic compass},
doi = {https://doi.org/10.1002/lpor.202400465},
url = {https://onlinelibrary.wiley.com/doi/abs/10.1002/lpor.202400465},
abstract = {Abstract The relationship between the magnetic field direction and the spatial intensity distribution of a radially polarized light passing through a polarized thermal atom ensemble is investigated, which is intuitively presented in a polarization selection absorption effect of thermal atoms. The radially polarized light has a spatial axisymmetric polarization structure, which is set as the probe beam. If the direction of the applied magnetic field is transformed, the absorption of the alignment atomic system to special polarization components of the probe light is changed, resulting in a different absorption ratio. This allows the 3D vector direction of the magnetic field to be inferred by using only the absorption ratio and the projection coefficient of the transmission intensity pattern. Based on this, this work provides a compass based on a thermal atom system, demonstrating a new method for measuring the magnetic field direction in space.},
year = {2024}
}

@article{Ramakrishna/PRA:2024,
  title = {Interaction of vector light beams with atoms exposed to a time-dependent magnetic field},
  author = {Ramakrishna, S. and Schmidt, R. P. and Peshkov, A. A. and Franke-Arnold, S. and Surzhykov, A. and Fritzsche, S.},
  journal = {Phys. Rev. A},
  volume = {110},
  issue = {4},
  pages = {043101},
  numpages = {10},
  year = {2024},
  month = {Oct},
  publisher = {American Physical Society},
  doi = {10.1103/PhysRevA.110.043101},
  url = {https://link.aps.org/doi/10.1103/PhysRevA.110.043101}
}

@article{Jones/AJP:16,
    author = {Jones, Joshua A. and D’Addario, Anthony J. and Rojec, Brett L. and Milione, G. and Galvez, Enrique J.},
    title = {The Poincaré-sphere approach to polarization: Formalism and new labs with Poincaré beams},
    journal = {American Journal of Physics},
    volume = {84},
    number = {11},
    pages = {822-835},
    year = {2016},
    month = {11},
    abstract = {We present a geometric-analytic introductory treatment of polarization based on the circular polarization basis, which connects directly to the Poincaré sphere. This treatment enables a more intuitive way to arrive at the polarization ellipse from the components of the field. We also present an advanced optics lab that uses Poincaré beams, which have a polarization that is spatially variable. The physics of this lab can reinforce understanding of all states of polarization, and in particular, elliptical polarization. In addition, it exposes students to Laguerre-Gauss modes, the spatial modes used in creating Poincaré beams, which have unique physical properties. In performing this lab, students gain experience in experimental optics, such as aligning and calibrating optical components, using and programming a spatial light modulator, building an interferometer, and performing polarimetry measurements. We present the apparatus for doing the experiments, detailed alignment instructions, and lower-cost alternatives.},
    issn = {0002-9505},
    doi = {10.1119/1.4960468},
    url = {https://doi.org/10.1119/1.4960468},
}

@article{Li/JAP:19,
    author = {Li, Delin and Feng, Shaotong and Nie, Shouping and Chang, Chenliang and Ma, Jun and Yuan, Caojin},
    title = {Generation of arbitrary perfect Poincaré beams},
    journal = {Journal of Applied Physics},
    volume = {125},
    number = {7},
    pages = {073105},
    year = {2019},
    month = {02},
    abstract = {},
    issn = {0021-8979},
    doi = {10.1063/1.5079850},
    url = {https://doi.org/10.1063/1.5079850},
}

@article{Beckley/OE:10,
author = {Amber M. Beckley and Thomas G. Brown and Miguel A. Alonso},
journal = {Opt. Express},
keywords = {Birefringence; Polarization; Azimuthally polarized beams; Cylindrical vector beams; Graded index fibers; Light beams; Optical fields; Step index fibers},
number = {10},
pages = {10777--10785},
publisher = {Optica Publishing Group},
title = {Full Poincar\'{e} beams},
volume = {18},
month = {May},
year = {2010},
url = {https://opg.optica.org/oe/abstract.cfm?URI=oe-18-10-10777},
doi = {10.1364/OE.18.010777},
abstract = {},
}

@article{Tian/PR:24,
author = {Ke Tian and Weifeng Ding and Zhaoying Wang},
journal = {Photon. Res.},
keywords = {Azimuthally polarized beams; Circular polarization; Optical tweezers; Spatial light modulators; Tunable diode lasers; Vector beams},
number = {5},
pages = {1093--1097},
publisher = {Optica Publishing Group},
title = {Dead-zone-free atomic magnetometer based on hybrid Poincar\&\#x00E9; beams},
volume = {12},
month = {May},
year = {2024},
url = {https://opg.optica.org/prj/abstract.cfm?URI=prj-12-5-1093},
doi = {10.1364/PRJ.519409},
abstract = {}
}

@article{Zhao/OLT:23,
title = {High-sensitivity pump–probe atomic magnetometer based on single fiber-coupled},
journal = {Optics \& Laser Technology},
volume = {159},
pages = {109025},
year = {2023},
issn = {0030-3992},
doi = {https://doi.org/10.1016/j.optlastec.2022.109025},
url = {https://www.sciencedirect.com/science/article/pii/S0030399222011719},
author = {Binbin Zhao and Junjian Tang and Hongying Yang and Lin Li and Yaohua Zhang and Ying Liu and Yueyang Zhai},
keywords = {Spin-exchange relaxation-free, Optical-fiber coupling, Triaxial magnetic compensation, Atomic magnetometer},
abstract = {}
}

@article{Shah/NP:07,
author={Shah, Vishal
and Knappe, Svenja
and Schwindt, Peter D. D.
and Kitching, John},
title={Subpicotesla atomic magnetometry with a microfabricated vapour cell},
journal={Nature Photonics},
year={2007},
month={Nov},
day={01},
volume={1},
number={11},
pages={649-652},
abstract={},
issn={1749-4893},
doi={10.1038/nphoton.2007.201},
url={https://doi.org/10.1038/nphoton.2007.201}
}

@article{Savukov/S:16,
AUTHOR = {Savukov, Igor and Boshier, Malcolm G.},
TITLE = {A High-Sensitivity Tunable Two-Beam Fiber-Coupled High-Density Magnetometer with Laser Heating},
JOURNAL = {Sensors},
VOLUME = {16},
YEAR = {2016},
NUMBER = {10},
ARTICLE-NUMBER = {1691},
URL = {https://www.mdpi.com/1424-8220/16/10/1691},
PubMedID = {27754358},
ISSN = {1424-8220},
ABSTRACT = {},
DOI = {10.3390/s16101691}
}

@article{Lange/PRL:22,
  title = {Excitation of an Electric Octupole Transition by Twisted Light},
  author = {Lange, R. and Huntemann, N. and Peshkov, A. A. and Surzhykov, A. and Peik, E.},
  journal = {Phys. Rev. Lett.},
  volume = {129},
  issue = {25},
  pages = {253901},
  numpages = {5},
  year = {2022},
  month = {Dec},
  publisher = {American Physical Society},
  doi = {10.1103/PhysRevLett.129.253901},
  url = {https://link.aps.org/doi/10.1103/PhysRevLett.129.253901}
}

@article{Schulz/PRA:2020,
  title = {Generalized excitation of atomic multipole transitions by twisted light modes},
  author = {Schulz, S. A.-L. and Peshkov, A. A. and M\"uller, R. A. and Lange, R. and Huntemann, N. and Tamm, Chr. and Peik, E. and Surzhykov, A.},
  journal = {Phys. Rev. A},
  volume = {102},
  issue = {1},
  pages = {012812},
  numpages = {10},
  year = {2020},
  month = {Jul},
  publisher = {American Physical Society},
  doi = {10.1103/PhysRevA.102.012812},
  url = {https://link.aps.org/doi/10.1103/PhysRevA.102.012812}
}

@article{Matula/JPB:2013,
doi = {10.1088/0953-4075/46/20/205002},
url = {https://dx.doi.org/10.1088/0953-4075/46/20/205002},
year = {2013},
month = {oct},
publisher = {IOP Publishing},
volume = {46},
number = {20},
pages = {205002},
author = {O Matula and A G Hayrapetyan and V G Serbo and A Surzhykov and S Fritzsche},
title = {Atomic ionization of hydrogen-like ions by twisted photons: angular distribution of emitted electrons},
journal = {J. Phys. B},
}

@book{Blum:2012,
  title={Density Matrix Theory and Applications},
  author={Blum, Karl},
  year={2012},
  publisher={Springer, Berlin}
}

@book{Auzinsh:2010,
  title={Optically Polarized Atoms: Understanding Light-Atom Interactions},
  author={Auzinsh, Marcis and Budker, Dmitry and Rochester, Simon M},
  year={2010},
  publisher={Oxford University, Oxford}
}

@article{Wense:2020,
  title={The theory of direct laser excitation of nuclear transitions},
  author={von der Wense, Lars and Bilous, Pavlo V and Seiferle, Benedict and Stellmer, Simon and Weitenberg, Johannes and Thirolf, Peter G and P{\'a}lffy, Adriana and Kazakov, Georgy},
  journal={Eur. Phys. J. A},
  volume={56},
  pages={176},
  year={2020},
  publisher={Springer},
  url = {https://doi.org/10.1140/epja/s10050-020-00177-x}
}

@article{Tremblay/PRA:1990,
  title = {Optical pumping with two finite linewidth lasers},
  author = {Tremblay, P. and Jacques, C.},
  journal = {Phys. Rev. A},
  volume = {41},
  issue = {9},
  pages = {4989--4999},
  numpages = {0},
  year = {1990},
  month = {May},
  publisher = {American Physical Society},
  doi = {10.1103/PhysRevA.41.4989},
  url = {https://link.aps.org/doi/10.1103/PhysRevA.41.4989}
}

@article{Schmidt/PRA:2024,
  title = {Atomic photoexcitation as a tool for probing purity of twisted light modes},
  author = {Schmidt, R. P. and Ramakrishna, S. and Peshkov, A. A. and Huntemann, N. and Peik, E. and Fritzsche, S. and Surzhykov, A.},
  journal = {Phys. Rev. A},
  volume = {109},
  issue = {3},
  pages = {033103},
  numpages = {11},
  year = {2024},
  month = {Mar},
  publisher = {American Physical Society},
  doi = {10.1103/PhysRevA.109.033103},
  url = {https://link.aps.org/doi/10.1103/PhysRevA.109.033103}
}

@article{Fritzsche/CPC:2019,
  title={A fresh computational approach to atomic structures, processes and cascades},
  author={Stephan Fritzsche},
  journal={Comput. Phys. Commun.},
  volume={240},
  number={1},
  pages={1-14},
  year={2019},
  doi = {10.1016/j.cpc.2019.01.012},
}

@book{arfken2011/book,
  title={Mathematical methods for physicists: a comprehensive guide},
  author={Arfken, George B and Weber, Hans J and Harris, Frank E},
  year={2011},
  publisher={Academic press}
}

@book{Johnson:2007,
  title={Atomic Structure Theory},
  author={Johnson, Walter R},
  year={2007},
  publisher={Springer, New York}
}

@book{Rose:1957,
  title={Elementary Theory of Angular Momentum},
  author={Rose, Morris Edgar},
  year={1957},
  publisher={John Wiley \& Sons, New York}
}
\preprint{}
\title{Interaction of a Poincaré beam with optically polarized atoms in the presence of constant magnetic field}

\author{S.~Ramakrishna}

\email[]{shreyas.ramakrishna@uni-jena.de}
\affiliation{Helmholtz-Institut Jena, D-07743 Jena, Germany}%
\affiliation{GSI Helmholtzzentrum f\"ur Schwerionenforschung GmbH, D-64291 Darmstadt, Germany}
\affiliation{Theoretisch-Physikalisches Institut, Friedrich-Schiller-Universit\"at Jena, D-07743 Jena, Germany}

\author{S.~Fritzsche}
\affiliation{Helmholtz-Institut Jena, D-07743 Jena, Germany}%
\affiliation{GSI Helmholtzzentrum f\"ur Schwerionenforschung GmbH, D-64291 Darmstadt, Germany}
\affiliation{Theoretisch-Physikalisches Institut, Friedrich-Schiller-Universit\"at Jena, D-07743 Jena, Germany}

\date{\today}

\begin{abstract}
Recent studies have highlighted the frequent applications of structured light modes in optically pumped atomic magnetometers. In this work, we theoretically explore how a Poincaré beam probes an optically polarized atomic medium. Specifically, we consider atoms polarized by a plane wave with linear polarization, immersed in a constant external magnetic field. We analyze how the polarization of the pump and probe light fields, along with the external magnetic field, impact the absorption profile. To this end, we employ a density matrix approach based on the Liouville-von Neumann equation. Our results reveal that the absorption profile exhibits an asymmetric pattern that depends on the magnetic field strength and the mutual orientation of the pump and probe light propagation directions relative to the quantization axis. For illustration, we assume the incoming radiation drives an electric dipole transition, $5s \, ^2S_{1/2}$ ($F=1$) $\rightarrow$ $5p \, ^2P_{3/2}$ ($F=0$), in rubidium atoms subjected to a magnetic field. These findings may aid in designing future experiments on optically pumped atomic magnetometers utilizing structured light modes.
\end{abstract}

\newpage
\maketitle

\section{Introduction}\label{Sec.Intro}
Optically pumped atomic magnetometers can be used to detect magnetic fields by monitoring properties such as intensity or polarization of the light at room temperatures~\cite{Budker:2013}. This detection scheme has found significant applications in fields such as geophysics~\cite{Dang/APL:2010}, medicine~\cite{Johnson/PMB:2013}, and fundamental physics~\cite{Vasilakis/PRL:2009}. As a result, their use is steadily increasing compared to superconducting quantum interference devices, which need to to be operated at cryogenic temperatures~\cite{Budker/N:2007}. Additionally, notable progress has been made in developing compact and miniaturized atomic magnetometers~\cite{Shah/NP:2007}.

Traditional atomic magnetometers rely on light fields with uniform polarization across their beam cross-section. However, modern optical techniques have enabled the generation of light fields with spatially varying polarization profiles~\cite{Rubinsztein-Dunlop/JO:2017}. One prominent member of this class of light fields is known as vector light modes, for example: radially and azimuthally polarized beams. These light modes contain spatially variable \textit{linear} polarization states within their beam cross-section~\cite{Gbur:2017}. Moreover, vector light beams have been shown to excite locally varying magnetization profiles in atoms, which can be utilized for measuring both static and oscillating magnetic field components~\cite{Castellucci/PRL:2021,Qiu/PR:2021,Sun/OE:23,Cai/LPR:24,Ramakrishna/PRA:2024}. In addition, an evolved version of vector beams, known as Poincaré beams, can also be generated~\cite{Jones/AJP:16,Li/JAP:19}. These beams exhibit a richer polarization texture, meaning more than one state of polarization across their beam cross-section~\cite{Beckley/OE:10}. Beyond vector light beams, a recent experiment demonstrated the elimination of dead-zones in an atomic magnetometer successfully with the help of a Poincaré beam~\cite{Tian/PR:24}. 

Conventional atomic magnetometers operate in either single- or dual-beam configurations, with the latter offering higher sensitivity for magnetic field measurements~\cite{Shah/NP:07,Savukov/S:16,Zhao/OLT:23}. In atomic magnetometry using structured light, prior studies have employed both configurations. For example, Cai et al.~\cite{Cai/LPR:24} experimentally analyzed the effects of a constant magnetic field on the absorption profile of a radially polarized beam interacting with atomic vapor polarized by a linearly polarized plane wave. In their experiment, both pump and probe fields shared the same propagation axis. While coaxial pump-probe configurations facilitate miniaturization, practical challenges arise in filtering pump light prior to probe analysis. A simple solution is to use orthogonally propagating pump and probe beams. Here, we consider a pump beam propagating perpendicular to the probe direction. Specifically, we explore the use of a Full-Poincaré beam to probe the response of an atomic medium—polarized by a linearly polarized plane wave—to a constant magnetic field.

We first examine the interaction between a Poincaré beam and an unpolarized atomic target under a constant magnetic field. This foundation is then extended to analyze interactions with polarized atomic targets. Our analysis reveals that the Poincaré beam's absorption profile exhibits axial asymmetry, contrasting with the symmetric profile observed for vector beams. Furthermore, this asymmetric absorption depends on the relative orientation between the pump/probe propagation directions and the quantization axis. We also investigate how the asymmetry varies with magnetic field strength, introducing an asymmetry parameter to quantify this relationship. This parameter provides a means to determine magnetic field strength. To demonstrate these principles, we model the electric dipole transition $5s,{}^2S_{1/2}$ ($F=1$) $\rightarrow$ $5p,{}^2P_{3/2}$ ($F=0$) in Rb atoms subjected to a constant magnetic field. These findings offer guidance for future experiments with optically pumped atomic magnetometers using structured light modes.

This paper is structured as follows: A brief mathematical description of a linearly polarized plane wave and Poincaré beam in Bessel basis is provided in Sec.~\ref{subsec:pumpprobe}. In Sec.~\ref{subsec:transition}, we derive the required transition amplitude of the interaction between the pump and probe light field with the atomic target. To determine the effect of the applied magnetic field on
the absorption profile of a Poincaré beam, we employ density matrix theory, whose basic formulas are briefly reviewed in Sec.~\ref{subsec:densitymatrices}. In Sec.~\ref{subsec:unpol}, we first discuss the absorption profile of the Poincaré beams interacting with unpolarized atomic target. As a next step, we discuss the absorption profile of Poincaré beams in the case of polarized atomic target, in Sec.~\ref{subsec:pol}. Finally, in Sec.~\ref{sec:summary} we provide a brief summary and outlook. Atomic units ($\hbar = m_{e} = e = 1, c = 1/\alpha$) are used throughout the manuscript.

\section{Theoretical background}

\subsection{Pump and Probe light fields}\label{subsec:pumpprobe}

\subsubsection{Plane waves}
In the current work, we consider the pump light field to be a linearly polarized plane wave. The vector potential of this plane wave, which is linearly polarized along the $x^{(\mathrm{pump})}$ axis, see Fig.~\ref{fig:geometry} and can be expressed as
\begin{equation}
     \bm{A}^{\mathrm{(lin)}}_{\mathrm{x}}(\bm{r}) = \frac{1}{\sqrt{2}} \left[ \bm{A}^{\mathrm{(circ)}}_{\lambda = +1}(\bm{r}) + \bm{A}^{\mathrm{(circ)}}_{\lambda = -1}(\bm{r}) \right],
\end{equation}
where $\bm{A}_{\lambda}^{(\mathrm{circ})} (\bm{r})$ is the vector potential of a circularly polarized plane wave with helicity $\lambda = \pm 1$, given by 
\begin{equation}
    \bm{A}_{\lambda}^{(\mathrm{circ})} (\bm{r}) = A_{0}\,\bm{e}_{\bm{k}\lambda}\; e^{i\bm{k}\cdot \bm{r}}.\label{eq:planevec}
\end{equation}
Here, $\bm{k}$ is the propagation vector, and $A_{0}$ is the amplitude, whose value will be specified later. 

\subsubsection{Poincaré beams}
\begin{figure}
    \centering
    \includegraphics[width=0.45\textwidth]{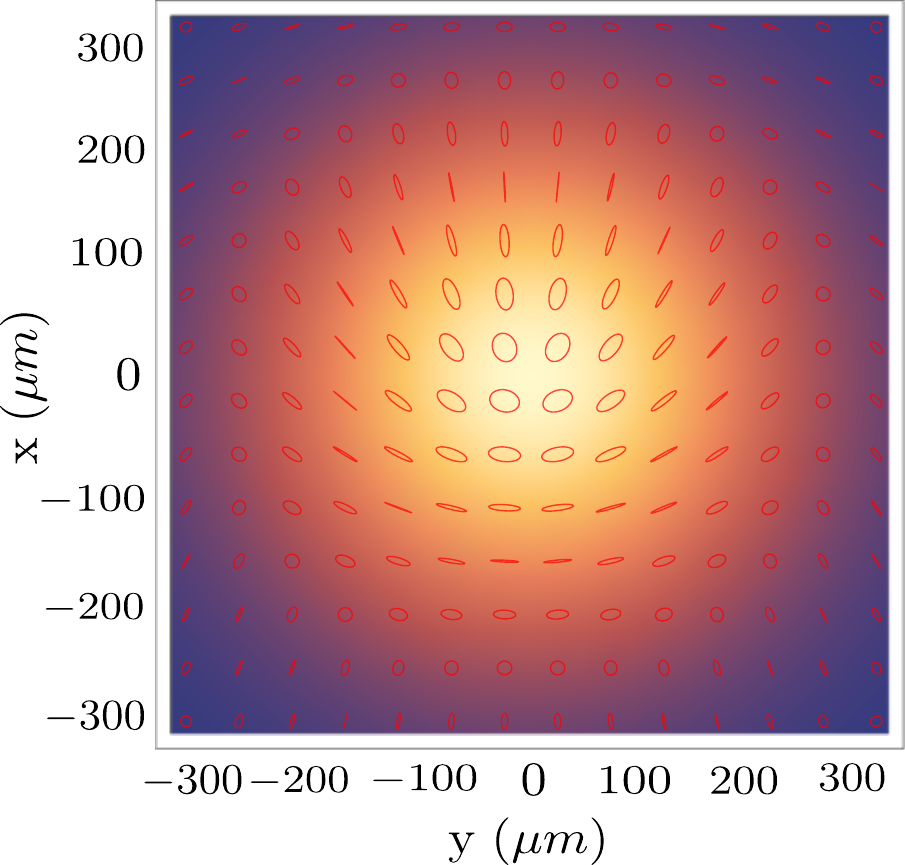}
    \caption{The intensity and polarization profile of a Poincaré beam constructed using the Bessel light of opening angle $\theta_{k} = 0.05\degree$.}
    \label{fig:Poincaré_int}
\end{figure}
We employ a Poincaré beam as the probe light field. Such beams are fundamentally constructed as superpositions of two or more circularly polarized structured light modes. Experimentally, this is typically realized using circularly polarized Laguerre-Gaussian (LG) modes. Theoretically, observed physical properties of these paraxial beams near their center can be accurately modeled using Bessel modes~\cite{Lange/PRL:22}. We therefore construct the Poincaré beam in the Bessel basis, with its vector potential expressed as
\begin{equation}
   \bm{A}^{\mathrm{(poin)}}(\bm{r}) =  \frac{1}{\sqrt{2}} \left[ \bm{A}^{\mathrm{(B)}}_{m_{\gamma} = +1, \, \lambda = +1}(\bm{r}) - \bm{A}^{\mathrm{(B)}}_{m_{\gamma} = 0 \, \lambda = -1}(\bm{r})  \right]. \label{eq:Poincarébessel}
\end{equation}

Here, $\bm{A}^{\mathrm{(B)}}_{m_{\gamma}, \, \lambda}(\bm{r})$ is the vector potential of circularly polarized Bessel light field carrying a projection of total angular momentum $m_{\gamma}$ onto its propagation axis. Since the theory of Bessel light fields has been frequently discussed in past publications~\cite{Matula/JPB:2013, Schulz/PRA:2020}, we will limit ourselves to the basic expressions. In particular, the vector potential of a circularly polarized Bessel light field can be written as
\begin{equation}
\bm{A}^{\mathrm{(B)}}_{m_{\gamma},\lambda}(\bm{r}) = A_{0}\int \frac{d^{2}\bm{k}_{\bot}}{(2\pi)^2} \; a_{\varkappa m_{\gamma}}(\bm{k}_{\bot}) \;\bm{e}_{\bm{k}\lambda} e^{i\bm{k}\cdot\bm{r}}  ,\label{eq:bessel_vec}
\end{equation}
\noindent where $a_{\varkappa m_{\gamma}}(\bm{k}_{\bot})$ is a weight function given by
\begin{equation}        
         a_{\varkappa m_{\gamma}}(\bm{k}_{\bot}) = \frac{2\pi}{\varkappa} (-i)^{m_{\gamma}} e^{im_{\gamma}\phi_{k}} \delta(k_{\bot}-\varkappa). \label{eq:bessel} 
\end{equation}
From the above expressions, one can understand Bessel light field as superposition of plane waves in momentum space whose wave vectors $\bm{k} = (k_{\bot},k_{z})$ lie on the surface of a cone with an opening angle of $\theta_{k} = \mathrm{arctan}(\varkappa/k_{z})$. By choosing smaller opening angle $\theta_{k}$, one can obtain paraxial Bessel light fields in which the transverse momentum is much weaker than its longitudinal counterpart, that is, $\varkappa \ll k_{z}$ (see Ref.~\cite{Schulz/PRA:2020}). By using this condition, one can approximate the vector potential of a Poincaré beam (\ref{eq:Poincarébessel}) as
\begin{align}
  \label{eq:poin}  \bm{A}^{\mathrm{(poin)}}(\bm{r},t) &\approx A_{0}\, \left[-i\,\left\{J_{0}(\varkappa r_{\bot}) + J_{1}(\varkappa r_{\bot})\, e^{i\phi_{r}}\right\}  \, \bm{e}_{x} \right.\\ \nonumber &\left. + \left\{J_{0}(\varkappa r_{\bot}) - J_{1}(\varkappa r_{\bot})\, e^{i\phi_{r}}\right\}\, \bm{e}_{y}  \right] \, e^{ik_{z}z}\, e^{i\omega t}. 
\end{align}
The electric field of the Poincaré beam is derived from its vector potential (Eq. \ref{eq:poin}) using $\bm{E}(\bm{r},t) = -\partial_{t}\bm{A}(\bm{r},t)$, enabling characterization of transverse intensity and polarization profiles. Although the intensity distribution resembles a Gaussian (Fig.~\ref{fig:Poincaré_int}), the beam exhibits a complex polarization structure: The local state evolves continuously from pure circular polarization at the beam center to elliptical with position-dependent ellipticity radially outward. Crucially, while the ellipticity becomes small far from the center, the polarization never collapses into a perfectly linear state. This residual ellipticity may appear linear in Fig.~\ref{fig:Poincaré_int} due to scale limitations. Our implementation combines structured light modes with orbital angular momentum projections $m_{\ell}=0$ ($m_{\gamma}=+1$) and $m_{\ell}=1$ ($m_{\gamma}=0$), though other combinations are possible. This configuration—termed a Full-Poincaré beam—generates a lemon-shaped polarization pattern~\cite{Jones/AJP:16}.

\subsection{Transition amplitudes}\label{subsec:transition}
\begin{figure*}
    \centering
    \includegraphics[width=0.7\textwidth]{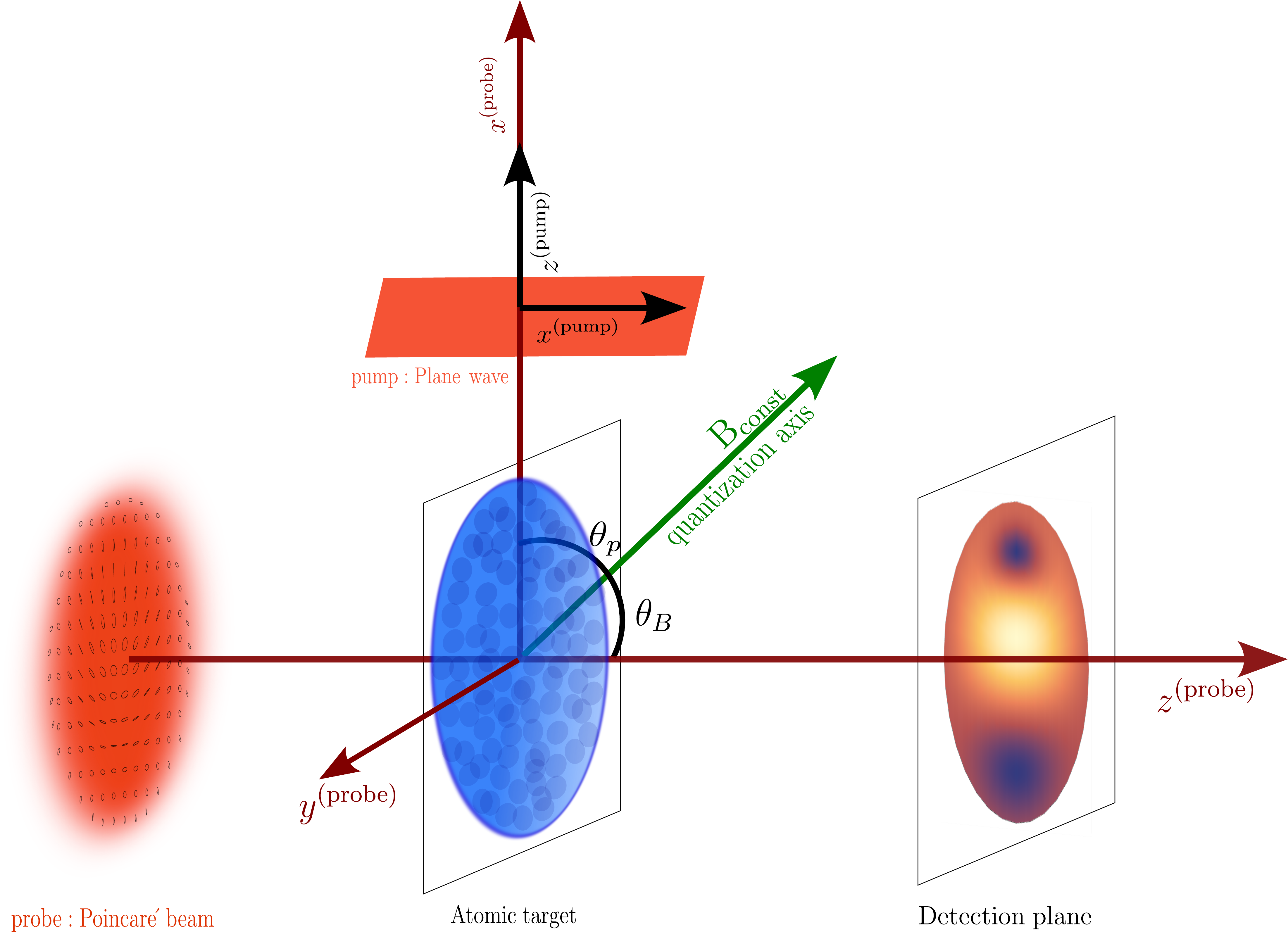}
    \caption{Geometry of the proposed pump-probe atomic magnetometer setup. The atomic target interacts with a pump and a probe light propagating perpendicular to each other along $z^{(\mathrm{pump})}$ and $z^{(\mathrm{probe})}$, respectively. The pump, plane wave is linearly polarized along the $x^{(\mathrm{pump})}$ direction and the probe, Poincaré beam has complex polarization texture. The external magnetic field, $\bm{B}_{\mathrm{const}}$, is applied at an angle of $\theta_{B}$ with respect to the probe light field. The quantization axis of this system is chosen along the applied constant magnetic field.}
    \label{fig:geometry}
\end{figure*}
Following the discussion of the mathematical formulation of pump and probe light fields, we now proceed to derive transition amplitude for the light atom interaction process. In particular, we will consider atomic transition from the initial $\vert \alpha_{g}F_{g}M_{g}\rangle$ to final $\vert \alpha_{e}F_{e}M_{e} \rangle$ atomic state driven by the incoming laser in the presence of external magnetic field $\bm{B}_{\mathrm{(const)}} = (0,0,B_{\mathrm{(const)}})$. Moreover, this external constant magnetic field is chosen as the quantization axis of the total system through out this paper. Then, one could write the first order transition matrix element as
\begin{align}
   V_{eg} = \, \frac{1}{\alpha} \; \left\langle \alpha_{e}F_{e}M_{e} \left \vert \sum_{q} \bm{\alpha}_{q} \cdot \bm{A}(\bm{r}_{q})\right \vert \alpha_{g} F_{g} M_{g} \right \rangle, \label{eq:amp}
\end{align}

where $\alpha$ is the fine structure constant, $\boldsymbol{F} = \boldsymbol{I} + \boldsymbol{J}$, where $\boldsymbol{I}$ and $\boldsymbol{J}$ are the nuclear and electron angular momenta, respectively, $M$ is the projection of $\boldsymbol{F}$ on the quantization axis, $\alpha_{g} (\alpha_{e})$ denotes all additional quantum numbers required to specify the ground (and excited) state uniquely. Moreover, $q$ runs over all electrons in a target atom and $\bm{\alpha}_q$ denotes the vector of Dirac matrices for the $q$th particle \cite{Johnson:2007}. In the above transition amplitude (\ref{eq:amp}), $\bm{A}(\bm{r})$ denotes the vector potential of either pump or probe light field. Before proceeding further, it can be noted that the transition matrix elements for the interaction process between linearly polarized plane wave, and a Poincaré beam with atom can be constructed in a similar way to their respective vector potentials. 
\begin{widetext}
The transition amplitudes for a circularly polarized plane wave of helicity $\lambda = \pm 1$ is given by
\begin{align}
\label{eq:pTME}
    V_{eg}^{(\mathrm{circ})}(\lambda) &= \frac{A_{0}}{\alpha} \sqrt{2\pi} \; \sum_{L,p} i^{L} \; [L,F_{g}]^{1/2} \; (i\lambda)^{p}\,  D^{L}_{\Delta M,\lambda}(-\pi,\theta_{p},-\pi)  \langle F_{g} M_{g}, L \Delta M| F_{e} M_{e}\rangle  \\ \nonumber & \times (-1)^{I+F_{g}+L+J_{e}} \, 
    \begin{Bmatrix}
        F_e & F_g & L \\
        J_g & J_e & I 
    \end{Bmatrix}\, \langle \alpha_{e}J_{e}|| \sum_{q} \bm{\alpha}_{q} \cdot \bm{a}_{L,q}^{(p)}||\alpha_{g} J_{g}\rangle,
\end{align}
and the transition amplitude for a Bessel light field with a given $m_{\gamma}$ and $\lambda = \pm1$ can be written as
\begin{align}
\label{eq:TME}    
V_{eg}^{(\mathrm{B})}(m_{\gamma},\lambda) &= \frac{A_{0}}{\alpha} \sqrt{2\pi} \; \sum_{M} i^{L+M}  \; [L,F_{g}]^{1/2} \; (i\lambda)^{p}\, (-1)^{m_\gamma} \; e^{i(m_\gamma-M)\phi_{b}} \; J_{m_{\gamma}-M}(\varkappa b) \, d^{L}_{M,\lambda}(\theta_{k}) \, \\     
& \times D^{L}_{\Delta M,M}(\pi,\theta_{B},\pi) \, \langle F_{g} M_{g}, L \Delta M| F_{e} M_{e}\rangle  \; (-1)^{I+F_{g}+L+J_{e}} \, \notag \\
    & \times
    \begin{Bmatrix}
        F_e & F_g & L \\
        J_g & J_e & I 
    \end{Bmatrix}
   \langle \alpha_{e}J_{e}|| \sum_{q} \bm{\alpha}_{q} \cdot \bm{a}_{L,q}^{(p)}||\alpha_{g} J_{g}\rangle. \notag
\end{align}

\end{widetext}
In the above expressions, $\Delta M = M_{e} - M_{g}$, $[L,F_{g}] = (2L+1)(2F_{g}+1)$, $p = 1$ (or 0) denotes electric (or magnetic) atomic transition with a multipolarity $L$, $\langle \alpha_{e}J_{e}|| \sum_{q} \bm{\alpha}_{q} \cdot \bm{a}_{L,q}^{(p)}||\alpha_{g} J_{g}\rangle$ is the reduced matrix elements, and $D^{L}_{\Delta M,\lambda}(\pi,\theta_{B},\pi)$ is the Wigner D functions. Here, the arguments of the Wigner D functions are Euler angles that characterize the rotation from the atomic frame with the quantization axis along the magnetic field $\bm{B}_{\mathrm{const}}$ to the photon frame with the quantization axis along the respective wave vectors $\bm{k}$ \cite{Rose:1957}. Moreover, the angle $\theta_{p} = \pi/2 -\theta_{B}$ ensures that at all the given times the pump and probe light fields are propagating perpendicular to each other, as shown in Fig.~\ref{fig:geometry}. Furthermore, taking into account the fact that Poincaré beam has a complex spatial structure, we have introduced impact parameter $\bm{b} = (b \cos \phi_b, b \sin \phi_b, 0)$ in (\ref{eq:TME}) which will characterize the position of the atom within its beam cross-section~\cite{Schulz/PRA:2020}. 

One can write the transition amplitude for linearly polarized pump light field interacting with atomic target as
\begin{align}
V_{eg,x}^{\mathrm{(lin)}} = \, \frac{1}{\sqrt{2}} \, \left[V_{eg}^{(\mathrm{circ})}(\lambda = +1) + V^{(\mathrm{circ})}_{eg}( \lambda = -1) \right]. \label{eq:linpln}
\end{align}    
Similarly, the transition amplitude for the probe Poincaré beam can be written as
\begin{align}
V_{eg}^{\mathrm{(poin)}} &= \, \frac{1}{\sqrt{2}} \, \left[V_{eg}^{(\mathrm{B})}(m_{\gamma} = +1,\lambda = +1)\right.\\ \nonumber &\left. - V^{(\mathrm{B})}_{eg}(m_{\gamma} = 0, \lambda = -1) \right]. 
\end{align}
It should be noted that, vector potential of (non-paraxial) Bessel light (\ref{eq:bessel_vec}) was used to derive the transition amplitude for the Poincaré beam atom interaction and this holds true for any arbitrary opening angle $\theta_{k}$. That being said, we will restrict ourselves to smaller $\theta_{k}$ values in our calculation to remain within paraxial regime.

\subsection{Density matrix formalism} \label{subsec:densitymatrices}
In this section, we model atomic state dynamics interacting with light fields using the Liouville-von Neumann equations~\cite{Blum:2012}. As this formalism was comprehensively detailed in our previous work~\cite{Ramakrishna/PRA:2024}, we present only essential expressions here. The atomic state is described by the density operator $\hat{\rho}(t)$, with its time evolution governed by
\begin{align}
    \frac{d}{dt}\hat{\rho}(t) = - i\, [\hat{H}(t),\hat{\rho}(t)] + \hat{R}(t).\label{eq:liouville}
\end{align}
In this expression, $\hat{H}(t)$ represents the total Hamiltonian of the system, encompassing interactions between the atomic system and the external magnetic field, pump field, and probe field. The relaxation operator $\hat{R}(t)$ phenomenologically models spontaneous decay from excited states to ground states~\cite{Auzinsh:2010}.

Following our previous approach~\cite{Ramakrishna/PRA:2024}, we construct the system's density matrix with dimensions $(2F_g + 2F_e + 2) \times (2F_g + 2F_e + 2)$. This formulation models transitions between atomic ground and excited states with multiple magnetic sublevels. Crucially, this formalism applies exclusively to two-level atomic systems. For such systems, the density matrix elements are expressed as:
\begin{subequations}
    \begin{align}
    \rho_{gg^{\prime}}(t) =& \langle \alpha_{g} F_{g} M_{g}| \hat{\rho}(t)| \alpha_{g} F_{g} M^{\prime}_{g}\rangle, \\ 
    \rho_{ee^{\prime}}(t) =& \langle \alpha_{e} F_{e} M_{e}| \hat{\rho}(t)| \alpha_{e} F_{e} M^{\prime}_{e}\rangle,  \\
    \rho_{ge}(t) =& \langle \alpha_{g} F_{g} M_{g}| \hat{\rho}(t)| \alpha_{e} F_{e}M_{e}\rangle, \\
    \rho_{eg}(t) =& \langle \alpha_{e} F_{e} M_{e}| \hat{\rho}(t)| \alpha_{g} F_{g}M_{g}\rangle.
    \end{align}\label{eq:elements}
\end{subequations}

Because the Liouville-von Neumann equations adopt a matrix representation in our formalism, the density matrix elements can be explicitly expressed within the rotating-wave approximation~\cite{Auzinsh:2010, Wense:2020} as:
\begin{widetext}
 \begin{subequations}
 \allowdisplaybreaks
    \begin{align}
    \allowdisplaybreaks
    \frac{d}{dt}\tilde{ \rho}_{g g^\prime}(t;M_{g},M_{e}) =& - i \Omega_g^{\mathrm{(L)}}  \left[M_{g} - M^{\prime}_{g} \right] \tilde{\rho}_{g g^\prime}(t) \; - \frac{i}{2} \left[ \sum_{M_e} V_{e g}^{*} \, \tilde{\rho}_{e g^\prime}(t) - \sum_{M_e} V_{e g^\prime} \, \tilde{\rho}_{g e}(t) \right] + R_{g g^\prime}(t) ,\\
    \frac{d}{dt}\tilde{\rho}_{e e^\prime}(t;M_{g},M_{e}) = &\, - i \Omega_e^{\mathrm{(L)}}  \left[M_{e} - M^{\prime}_{e} \right] \tilde{\rho}_{e e^\prime}(t) \; - \frac{i}{2} \left[ \sum_{M_g} V_{e g} \, \tilde{\rho}_{g e^\prime}(t) - \sum_{M_g} V_{e^\prime g}^{*} \, \tilde{\rho}_{e g}(t) \right] + R_{e e^\prime}(t) ,\\
    \frac{d}{dt} \tilde{\rho}_{g e}(t;M_{g},M_{e}) =& \; -i \Delta \tilde{\rho}_{g e}(t) + i \left[ \Omega_e^{\mathrm{(L)}}  M_{e} - \Omega_g^{\mathrm{(L)}}  M_{g} \right] \tilde{\rho}_{g e}(t) \; - \frac{i}{2} \left[ \sum_{M_e^\prime} V_{e^\prime g}^{*} \, \tilde{\rho}_{e^\prime e}(t) - \sum_{M_g^\prime} V_{e g^\prime}^{*} \, \tilde{\rho}_{g g^\prime}(t) \right] + R_{g e}(t) ,\\
   \frac{d}{dt} \tilde{\rho}_{eg}(t;M_{g},M_{e})  =& \; i \Delta \tilde{\rho}_{e g}(t) - i \left[ \Omega_e^{\mathrm{(L)}}  M_{e} - \Omega_g^{\mathrm{(L)}}  M_{g} \right] \tilde{\rho}_{e g}(t) \; - \frac{i}{2} \left[ \sum_{M_g^\prime} V_{e g^\prime} \; \tilde{\rho}_{g^\prime g}(t) - \sum_{M_e^\prime} V_{e^\prime g} \; \tilde{\rho}_{e e^\prime}(t) \right] + R_{e g}(t).
    \end{align}\label{eq:diff-eqn}
 \end{subequations}
\end{widetext}

Here, $\Delta$ denotes the light frequency detuning from resonance, and $\Omega^{\mathrm{(L)}} = g_{F} \mu_{B} B_{\mathrm{const}} /\hbar$ is the Larmor frequency. We also assume the pump and probe fields share the same angular frequency, thereby driving transitions between identical hyperfine levels. The total transition amplitude $V_{eg}$ is given by $V_{eg} = V_{eg}^{\mathrm{(pump)}} + V_{eg}^{\mathrm{(probe)}}$, representing the sum of matrix elements for pump-atom and probe-atom interactions.  

Finally, the spontaneous decay $R(t)$ terms required to solve the above Liouville-von Neumann equations is expressed as 
\begin{subequations}
\allowdisplaybreaks
\begin{align}
\allowdisplaybreaks
    \nonumber &R_{g g^\prime} (t) = \;\Gamma [F_{g},J_{e}] \begin{Bmatrix}
F_e & F_g & L \\
J_g & J_e & I
\end{Bmatrix}^2 \\
    &\times \sum\limits_{M_{e}, M^{\prime}_{e}, M} \langle F_g M_{g} L M | F_e M_{e} \rangle \, \tilde{\rho}_{e e^\prime} (t) \, \langle F_g M^\prime_{g} L M | F_e M^\prime_{e} \rangle  \, , \\
    &R_{e e^\prime} (t) = \, - \Gamma [F_{g},J_{e}] \begin{Bmatrix}
F_e & F_g & L \\
J_g & J_e & I
\end{Bmatrix}^2 \tilde{\rho}_{e e^\prime} (t), \\
    &R_{g e} (t) = \, - \frac{1}{2} \Gamma [F_{g},J_{e}] \begin{Bmatrix}
F_e & F_g & L \\
J_g & J_e & I
\end{Bmatrix}^2 \tilde{\rho}_{g e} (t) \, , \\
    &R_{e g} (t) = \, - \frac{1}{2} \Gamma [F_{g},J_{e}] \begin{Bmatrix}
F_e & F_g & L \\
J_g & J_e & I
\end{Bmatrix}^2 \tilde{\rho}_{e g} (t),  
\end{align}
\label{eq:RTerms}
\end{subequations}

\noindent where $\Gamma$ is the decay rate of the upper level $|\alpha_{e} J_{e}\rangle$ \cite{Tremblay/PRA:1990, Schmidt/PRA:2024,Ramakrishna/PRA:2024}.

\section{Results and Discussion}
In the present work, we investigate the interaction between a Poincaré beam and atoms in the presence of an external magnetic field. In experiments related to atomic magnetometers, this interaction between the probe light and atoms is quantified by detecting the absorption profile of the transmitted light~\cite{Castellucci/PRL:2021,Cai/LPR:24}. Specifically, the absorption profile reveals which parts of the beam cross-section were absorbed by the interacting atoms. Theoretically, this can be examined by monitoring the populations of the excited and ground states of the atoms across the beam cross-section. This approach was outlined in our earlier publication~\cite{Ramakrishna/PRA:2024} and is equivalent to calculating the imaginary part of the refractive index of the polarized atomic medium.

The theoretical framework developed in the previous section applies to atomic transitions between any two hyperfine levels. Here, we consider a rubidium ensemble with the ground state \(5S_{1/2}\,(F_g = 1)\) and excited state \(5P_{3/2}\,(F_e = 0)\), focusing on the \(F_g = 1 \rightarrow F_e = 0\) transition at the resonance frequency \(\omega_0 = 2\pi \times 384\,\text{THz}\). We neglect thermal motion and Doppler broadening, which would otherwise require taking into account all excited-state hyperfine levels. Both the pump and probe fields are assumed to drive this transition simultaneously. Furthermore, interactions between the two lasers—such as electromagnetically induced transparency—are neglected in our analysis. This is because the probe light couples multiple ground-state magnetic sublevels (\(m_g = -1, 0, +1\)) to the excited state. As a result, the pump and probe do not selectively address two distinct sublevels, and the system does not fulfill the necessary conditions for coherent interactions between the two lasers.

Field amplitudes are set to \(A_0^{\mathrm{(pump)}} = 1.28 \times 10^{-13}\) and \(A_0^{\mathrm{(probe)}} = 2.02 \times 10^{-14}\) with a Bessel mode opening angle \(\theta_k = 0.05^\circ\). These parameters emulate experimental Laguerre-Gaussian conditions: a 300 $\mu$m beam waist for both Gaussian (pump) and Poincar\'{e} (probe) beams. The pump power exceeds the probe by a factor of ten, consistent with standard magnetometer configurations~\cite{Cai/LPR:24}.

Along with these beam parameters, we require the spontaneous decay rate $\Gamma$ and the reduced matrix element $\langle \alpha_{e}J_{e} \| \sum_{q} \bm{\alpha}_{q} \cdot \bm{a}_{L,q}^{(p)} \| \alpha_{g} J_{g} \rangle$. The decay rate $\Gamma = 4.042 \times 10^7~\text{s}^{-1}$ is obtained from the Jena Atomic Calculator (JAC) code~\cite{Fritzsche/CPC:2019}, which also enables calculation of the reduced matrix element. These parameters determine the transition amplitude $V_{eg}$ for both pump and probe interactions with the rubidium target.

The steady-state absorption profile of the Poincar\'{e} beam is obtained by computing the excited-state population distribution across its beam cross-section. This requires solving the Liouville-von Neumann equations (\ref{eq:diff-eqn}) in steady state ($\partial\hat{\rho}/\partial t = 0$), resulting in a system of 16 coupled linear equations. Due to linear dependence among these equations, only 10 independent equations need solving, which we implement using the computer algebra system \textit{Mathematica}.

\subsection{Unpolarized atomic target}\label{subsec:unpol}

\begin{figure}
    \centering
    \includegraphics[width=0.5\textwidth]{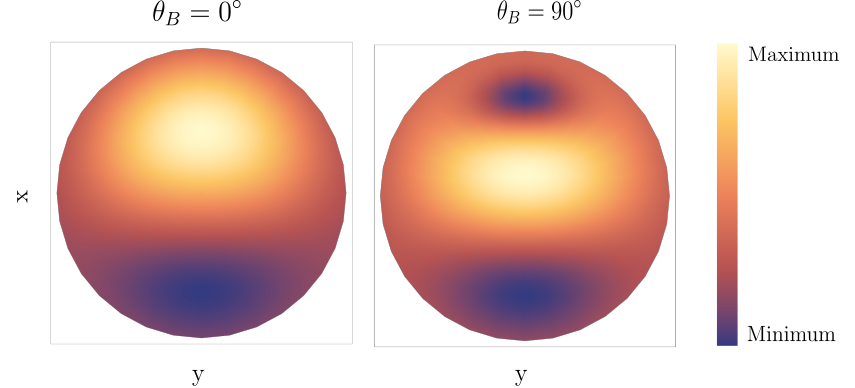}
    \caption{The asymmetrical absorption profile of the Poincaré beam interacting with an unpolarized rubidium atomic target immersed in an external constant magnetic field of strength \( B_{\mathrm{(const)}} = 1 \, \mathrm{G} \). The absorption profile is shown for the Poincaré beam propagating at an angle of (left) \( \theta_{B} = 0^\circ \) and (right) \( 90^\circ \) with respect to the quantization axis. In these absorption profiles, the bright orange color corresponds to the region in the beam cross-section where the Poincaré beam is maximally absorbed by the atoms, while the dark purple color represents the region with minimal absorption.}
    \label{fig:absorption_nopump}
\end{figure}

Let us begin our discussion by considering the interaction of a Poincaré beam with an unpolarized atomic target, meaning the pump light field is turned off. In this simplified scenario, all three magnetic sublevels in the ground state are assumed to be equally populated. For this case, we examine the population of the excited atomic state across the beam cross-section under steady-state conditions to construct the absorption profile. Moreover, we assume the strength of the external magnetic field to be \(B_{\mathrm{(const)}} = 1\, \mathrm{G}\). For clarity, we focus on two cases: (a) Poincaré probe field propagating parallel (\(\theta_{B} = 0^\circ\)) and (b) perpendicular (\(90^\circ\)) to the quantization axis. As shown in Fig.~\ref{fig:absorption_nopump}, the absorption profile exhibits asymmetry along the \(y\)-axis in both cases. In the following, we will provide a detailed explanation to understand these absorption profiles.

\begin{figure}
    \centering
    \includegraphics[width=0.5\textwidth]{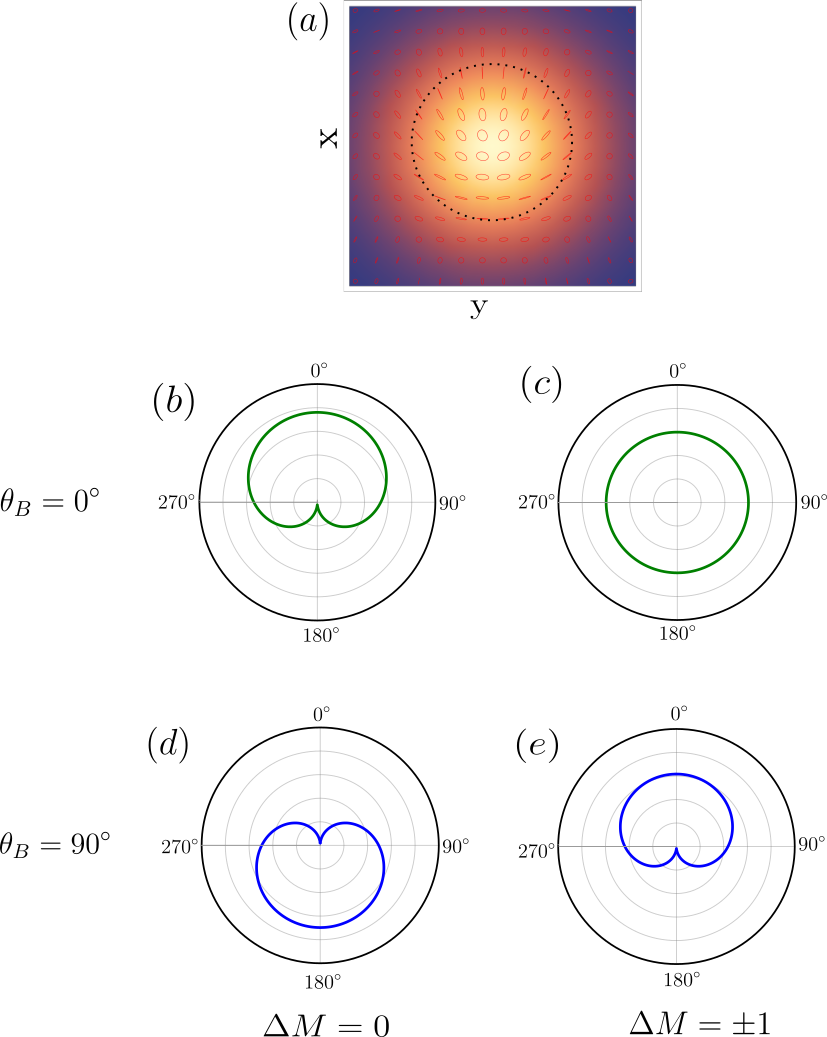}
    \caption{Polar plots of the absolute value of the transition amplitude \( |V_{eg}^{(\mathrm{poin})}| \) for the rubidium atom interacting with a Poincaré beam. (a) Here, \( |V_{eg}^{(\mathrm{poin})}| \) is calculated for a rubidium atom located on a circle with a radius of \( b = 200 \, \mu\mathrm{m} \), depicted by the dotted line on the Poincaré beam's cross-section. Along this circumference, the local polarization is found to be elliptical. In this case, the Poincaré beam propagates at an angle of \( \theta_{B} = 0^\circ \) and interacts with the atoms, coupling all three magnetic sublevels in the ground state to the excited state. The polar plot of the local \( |V_{eg}^{(\mathrm{poin})}| \), which corresponds to the circle of radius \( b \), is asymmetrical for the transition (b) \( M_{g} = 0 \rightarrow M_{e} = 0 \), and symmetrical for the transition (c) \( M_{g} = \pm 1 \rightarrow M_{e} = 0 \). When the Poincaré beam propagates at an angle of \( \theta_{B} = 90^\circ \), the polar plot of the local \( |V_{eg}^{(\mathrm{poin})}| \) corresponding to the radius \( b \) is asymmetrical for both transitions: (d) \( M_{g} = 0 \rightarrow M_{e} = 0 \), and (e) \( M_{g} = \pm 1 \rightarrow M_{e} = 0 \). Here, the transition amplitude is calculated by setting the following values for the beam parameters: \( \theta_{k} = 0.05^\circ \), \( A^{(\mathrm{probe})} = 2.02 \times 10^{-14} \).}
    \label{fig:transition_amp}
\end{figure}
\subsubsection{For $\theta_{B} = 0\degree$}
When $\theta_{B} = 0\degree$, Fig.~\ref{fig:absorption_nopump} reveals stronger light absorption by atoms in the upper beam cross-section, indicating enhanced interaction with the Poincar\'{e} beam. Conversely, atoms in the lower cross-section exhibit weaker interaction. This position-dependent variation reflects the beam's inhomogeneous polarization pattern, which manifests in the transition amplitude through the Bessel function $J_{m_{\gamma} - M}(\varkappa b)$. To understand the asymmetric absorption profile in Fig.~\ref{fig:absorption_nopump}, we analyze the local light-atom transition amplitude.  

In Fig.~\ref{fig:absorption_nopump}, as one moves away from the beam center along the positive $x$-axis, the population of atoms in the excited state increases. Specifically, the excited-state population reaches a maximum at $b = 200\,\mu\mathrm{m}$, $\phi_{b} = 0\degree$. In contrast, moving in the opposite direction (negative $x$-axis), the excited-state population decreases rapidly, attaining a minimum at $b = 200\,\mu\mathrm{m}$, $\phi_{b} = 180\degree$. This behavior can be explained in terms of the local polarization structure of the Poincaré beam and the corresponding transition amplitude $V_{eg}$. As shown in Fig.~\ref{fig:Poincaré_int}, the local polarization evolves from being purely circular at the center to elliptical as the radial distance increases. In simpler terms, the ellipticity of the polarization ellipse depends on the impact parameter $\bm{b}$. For instance, atoms located at $b = 200\,\mu\mathrm{m}$, $\phi_{b} = 0\degree$ experience a locally elliptically polarized light field, as shown in Fig.~\ref{fig:Poincaré_int}, where the major axis of the ellipse is aligned along the $x$-axis. On the other hand, atoms at $b = 200\,\mu\mathrm{m}$, $\phi_{b} = 180\degree$ encounter an elliptically polarized field with the major axis oriented along the $y$-axis, as illustrated in Fig.~\ref{fig:Poincaré_int}. It is important to note that the quantization axis in both cases remains perpendicular to the local polarization plane. As a result, atoms at these positions can undergo transitions from all three magnetic sublevels in the ground state ($M_{g} = 0, \pm 1$) to the excited state. However, the strengths of these individual transition amplitudes are not identical and depend sensitively on the atom's position within the beam cross-section.

In principle, the strength of the local transition amplitude $|V_{eg}|$ for atomic transitions between $M_{g} = \pm 1$ and $M_{e} = 0$ is found to be axially symmetric for any $b \neq 0$. As an example, Fig.~\ref{fig:transition_amp}(c) shows a polar plot of $|V_{eg}^{(\mathrm{poin})}|$ at a fixed radius of $b = 200\,\mu\mathrm{m}$. From this plot, we observe that the transition amplitude corresponding to $\Delta M = \pm 1$ ($M_{g} = \pm 1 \rightarrow M_{e} = 0$) is indeed axially symmetric. However, the transition amplitude for $M_{g} = 0 \rightarrow M_{e} = 0$ ($\Delta M = 0$) does not exhibit axial symmetry. This is shown in Fig.~\ref{fig:transition_amp}(b), where we present a polar plot of the absolute value of the $\Delta M = 0$ transition amplitude for $b = 200\,\mu\mathrm{m}$ as a function of the azimuthal angle $\phi_{b}$. These plots clearly reveal that the $\Delta M = 0$ transition amplitude lacks axial symmetry. This asymmetry arises from the inhomogeneous polarization profile of the Poincaré beam and is strongly position-dependent. In particular, the local transition amplitude for the $\Delta M = 0$ case attains its maximum at $b = 200\,\mu\mathrm{m}$, $\phi_{b} = 0\degree$ (upper half of the beam), and reaches a minimum at $b = 200\,\mu\mathrm{m}$, $\phi_{b} = 180\degree$ (lower half of the beam). As a result, atoms located in the lower part of the Poincaré beam cross-section exhibit weaker coupling with the light field compared to those in the upper part. This leads to the observed asymmetry in the absorption profile, where the atomic excited-state population is minimized in the lower half of the beam cross-section.

From Fig.~\ref{fig:absorption_nopump}, we observe that the atoms located at the center of the beam absorb relatively minimum light in comparison to those in the upper part of the beam cross-section. To understand this, we turn our attention to the local transition amplitude $V_{eg}$. Since the angle $\theta_{B} = 0\degree$, the Wigner-$D$ function in the transition amplitude expression~(\ref{eq:TME}) simplifies to a delta function: $D_{\Delta M, M}^{L}(\pi,0,\pi) = \delta_{\Delta M,M}$. Additionally, for atoms at the center ($\bm{b} = 0$), the Bessel function in the transition amplitude~(\ref{eq:TME}) reduces to $J_{m_{\gamma}-M}(0) = \delta_{m_{\gamma}-M,0}$. These simplifications yield the condition $\Delta M = M$, and the transition amplitude is non-zero only when $m_{\gamma} = M$. In our case, this implies that the transition amplitude is non-zero for $M = m_{\gamma} = +1$ and $M = m_{\gamma} = 0$. This is evident from the simplified expression for the transition amplitude $V_{eg}^{(\mathrm{poin})}$:
\begin{align}
    V_{eg}^{\mathrm{(poin)}} &\approx \, d^{1}_{m_{\gamma},1}(\theta_{k})\, \langle 1 M_{g}, 1 m_{\gamma}| 0 0\rangle \\ \nonumber
    &-\, d^{1}_{m_{\gamma},-1}(\theta_{k})\, \langle 1 M_{g}, 1 m_{\gamma}| 0 0\rangle,
\end{align}
where we have omitted other factors from the transition amplitude expression~(\ref{eq:TME}) as they do not influence the result. Consequently, two transitions are allowed: $M_{g} = -1 \rightarrow M_{e} = 0$ and $M_{g} = 0 \rightarrow M_{e} = 0$. However, since the local polarization of the Poincaré beam is circular at the center, the amplitude for the $M_{g} = -1 \rightarrow M_{e} = 0$ transition is found to be stronger than that for $M_{g} = 0 \rightarrow M_{e} = 0$. As the system evolves to a steady state, the population in the $M_{g} = -1$ sublevel depletes significantly compared to the other ground-state sublevels ($M_{g} = 0, +1$). In contrast, the $M_{g} = +1$ sublevel remains uncoupled to the light field and thus retains the highest population. Ultimately, the atom interacts only weakly with the light field via the residual $M_{g} = 0 \rightarrow M_{e} = 0$ transition. This results in a relatively lower excited-state population at the center of the beam cross-section compared to the upper part of the absorption profile, as shown in Fig.~\ref{fig:absorption_nopump}.

\subsubsection{For $\theta_{B} = 90\degree$}
Let us now consider the case where the Poincaré beam propagates perpendicular to the quantization axis, i.e., $\theta_{B} = 90^{\circ}$. In this scenario, the external magnetic field (which defines the quantization axis) lies along the $x$-axis. As discussed earlier, atoms located at $b = 200\, \mu\mathrm{m}$ and $\phi_{b} = 0^{\circ}$ experience a local elliptically polarized light field, as shown in Fig.~\ref{fig:Poincaré_int}(a). However, in this case, the major axis of the polarization ellipse is aligned with the quantization axis. Conversely, atoms located in the lower part of the beam, at $b = 200\, \mu\mathrm{m}$ and $\phi_{b} = 180^{\circ}$, encounter elliptically polarized light whose minor axis is aligned with the quantization axis. The calculation of the local transition amplitude in this configuration reveals that the absolute value of $|V_{eg}^{(\mathrm{poin})}|$ is asymmetric for both $\Delta M = 0$ and $\Delta M = \pm 1$ transitions, as illustrated in Figs.~\ref{fig:transition_amp}(d) and (e). Specifically, the transition amplitude $|V_{eg}^{(\mathrm{poin})}|$ for the $M_{g} = 0 \rightarrow M_{e} = 0$ transition attains its minimum in the upper part of the beam, particularly at $b = 200\, \mu\mathrm{m}$, $\phi_{b} = 0^{\circ}$. In contrast, the transition amplitude $|V_{eg}^{(\mathrm{poin})}|$ for $M_{g} = \pm 1 \rightarrow M_{e} = 0$ reaches its minimum in the lower part of the beam, especially at $b = 200\, \mu\mathrm{m}$, $\phi_{b} = 180^{\circ}$. As a result, atoms located in both the upper and lower halves of the Poincaré beam cross-section interact weakly with the light field. Notably, the transition amplitude $|V_{eg}^{(\mathrm{poin})}|$ is relatively smaller for atoms at $b = 200\, \mu\mathrm{m}$, $\phi_{b} = 180^{\circ}$ compared to those at $b = 200\, \mu\mathrm{m}$, $\phi_{b} = 0^{\circ}$. In conclusion, atoms positioned in the lower part of the beam cross-section absorb less light relative to those in the upper part. This leads to an asymmetric absorption profile featuring two dark lobes, as shown in Fig.~\ref{fig:absorption_nopump}.

In this scenario, with $\theta_{B} = 90^{\circ}$, atoms located at the beam center interact strongly with the incoming Poincaré beam—unlike the previous case with $\theta_{B} = 0^{\circ}$. Since $\bm{b} = 0$, the Bessel function in Eq.~(\ref{eq:TME}) simplifies to a delta function: $J_{m_{\gamma}-M}(0) = \delta_{m_{\gamma}-M,0}$. However, the Wigner-$D$ function is no longer a delta function in this configuration, and therefore $\Delta M \neq M$ in the general expression for the transition amplitude~(\ref{eq:TME}). As a result, transitions from all three magnetic sublevels of the ground state, $M_{g} = 0, \pm 1$, to the excited state $M_{e} = 0$ become allowed. Consequently, atoms at the center of the beam cross-section experience strong coupling to the light field. This leads to a relatively higher population of the excited state at $\bm{b} = 0$, compared to other regions of the beam cross-section.

\subsection{Polarized atomic target}\label{subsec:pol}
\begin{figure}
    \centering
    \includegraphics[width=0.48\textwidth]{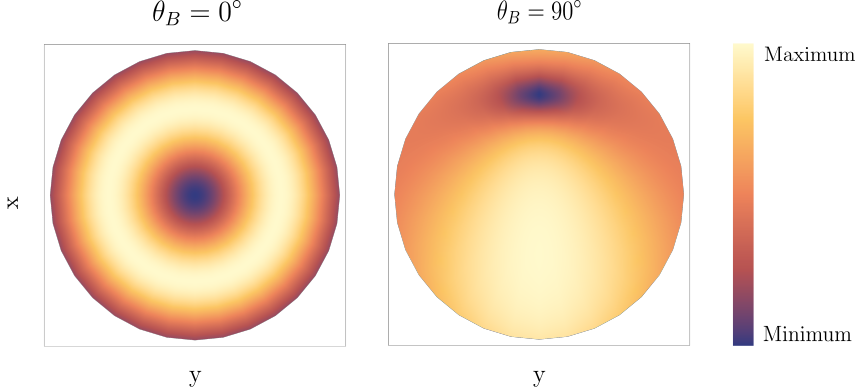}
    
    \caption{The absorption profile of the Poincaré beam interacting with a polarized rubidium atomic target immersed in an external constant magnetic field of strength \(B_{\mathrm{(const)}} = 1\, \mathrm{G}\). Here, the absorption profile is shown for the Poincaré beam propagating at an angle of (left) \(\theta_{B} = 0^\circ\) and (right) \(90^\circ\) with respect to the quantization axis. In these absorption profiles, the bright orange color corresponds to regions in the beam cross-section where the Poincaré beam is absorbed maximally by the atoms, and the dark purple color represents regions with minimal absorption.}
    \label{fig:absorption_linear}
\end{figure}
\begin{figure}
    \centering
    \includegraphics[width=0.45\textwidth]{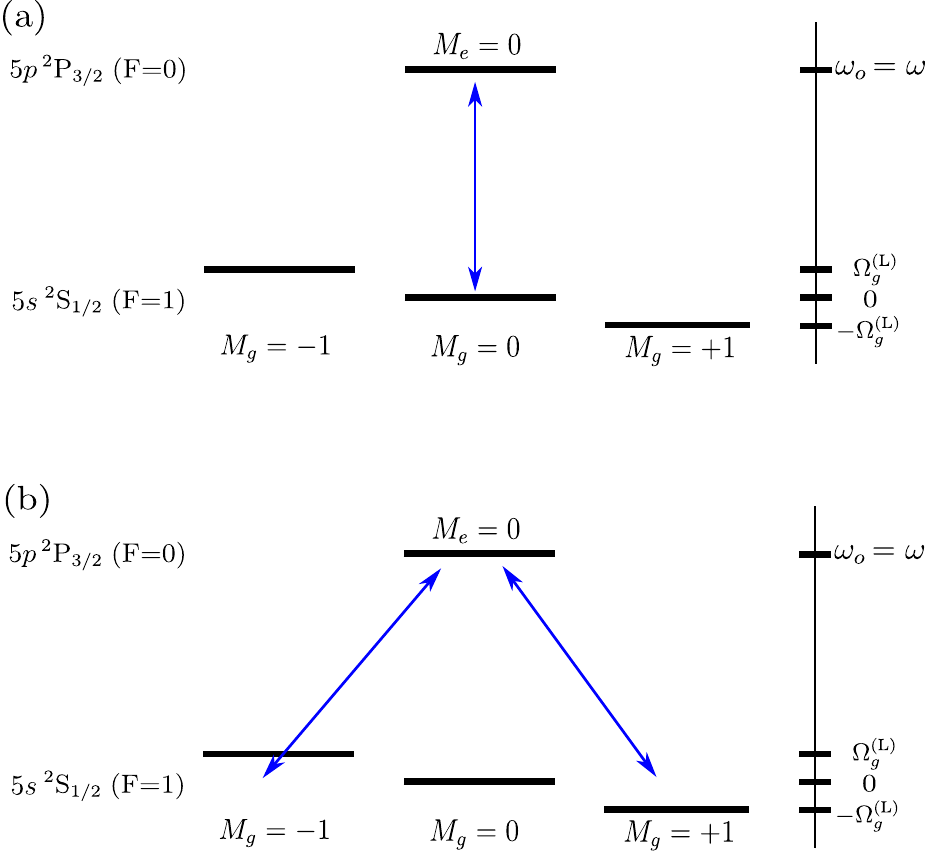}
    \caption{The transition between $5s \;\, {}^2 S_{1/2}$ ($F=1$) $-$ $5p \;\, {}^2 P_{3/2}$ ($F=0$) in $^{87}$Rb driven by the pump, linearly polarized plane wave for (a) $\theta_{B} = 0\degree$ and (b) $\theta_{B} = 90\degree$. The lower sublevels are split by the energy $\hbar \Omega_g^{\mathrm{(L)}}$ as given by the Larmor frequency of the atom in the magnetic field. }
    \label{fig:energy_level}
\end{figure}

We now examine the interaction between the incoming Poincaré beam and an optically polarized ensemble of rubidium atoms subjected to a constant magnetic field of strength $B_{\mathrm{(const)}} = 1$\,G. As in the previous case, we consider the Poincaré beam to propagate either parallel ($\theta_{B} = 0^{\circ}$) or perpendicular ($\theta_{B} = 90^{\circ}$) to the quantization axis. When $\theta_{B} = 0^{\circ}$, the pump light field propagates perpendicular to the quantization axis, resulting in the polarization of the pump field being aligned parallel to the quantization axis. In this configuration, the pump selectively drives atomic transitions satisfying the angular momentum selection rule $\Delta M = 0$ (see Fig.~\ref{fig:energy_level}(a)). As a consequence, the atomic medium becomes optically polarized, with the ground-state sublevel $M_{g} = 0$ being depopulated and the atomic population redistributed equally between the $M_{g} = \pm 1$ sublevels.
\subsubsection{For $\theta_{B} = 0\degree$}
In Fig.~\ref{fig:absorption_linear}, we present the steady-state absorption profile for the case where the Poincaré beam propagates at an angle $\theta_{B} = 0^{\circ}$. Under this configuration, the absorption pattern exhibits a characteristic \textit{donut-like} shape. Compared to the unpolarized case, two distinct features emerge: (a) the \textit{axial symmetry} of the absorption profile is restored, and (b) the excited-state population at the center of the beam cross-section ($\bm{b} = 0$) vanishes. In other words, atoms located at the beam center do not interact with the incoming Poincaré beam. To understand these features, we revisit the local transition amplitude $|V_{eg}^{(\mathrm{poin})}|$. As discussed earlier, for $\theta_{B} = 0^{\circ}$, the transition amplitude between the magnetic sublevels $M_{g} = 0$ and $M_{e} = 0$ is not axially symmetric (see Fig.~\ref{fig:transition_amp}(b)). However, since the atomic medium is polarized such that the population in the $M_{g} = 0$ ground-state sublevel is zero, transitions with $\Delta M = 0$ are effectively forbidden. Consequently, the Poincaré beam primarily induces transitions between the ground-state sublevels $M_{g} = \pm 1$ and the excited state $M_{e} = 0$ across the entire beam cross-section. Moreover, the absolute value of the transition amplitude $|V_{eg}^{(\mathrm{poin})}|$ for these $\Delta M = \pm 1$ transitions is axially symmetric, which explains the restored symmetry in the absorption profile. Therefore, the interaction of the Poincaré beam with a polarized atomic target at $\theta_{B} = 0^{\circ}$ results in an axially symmetric absorption profile.

Additionally, the absorption profile shows zero excited-state population for atoms located at the beam center, as shown in Fig.~\ref{fig:absorption_linear}. From our earlier discussion, we know that at the beam center, atoms experience circularly polarized light, and the transition amplitude is non-zero as long as $m_{\gamma} = M = 0$ or $+1$. That is, for $\bm{b} = 0$, transitions can occur between the magnetic sublevels $M_{g} = -1, 0$ and $M_{e} = 0$. However, in the present case of a polarized atomic target, the Poincaré beam drives only the transition from $M_{g} = -1$ to $M_{e} = 0$. As the system reaches steady state, only the magnetic sublevel $M_{g} = +1$ remains populated, while the other two sublevels are depleted. Consequently, the excited-state population at the beam center remains zero.

Thus, the incoming Poincaré beam is absorbed by atoms across the beam profile except those located at $\bm{b} = 0$, resulting in the observed \textit{donut-like} absorption profile in Fig.~\ref{fig:absorption_linear}.

\subsubsection{For $\theta_{B} = 90\degree$}
Now, we consider the case where the Poincaré beam propagates at an angle of $\theta_{B} = 90^\circ$ with respect to the quantization axis. In this configuration, the pump—a linearly polarized plane wave—propagates along the quantization axis, which lies in the transverse plane of the probe light field. Under these conditions, the pump light drives atomic transitions in rubidium that satisfy the selection rule $\Delta M = \pm 1$, as illustrated in Fig.~\ref{fig:energy_level}(b). Consequently, the population in the ground sublevel $M_{g} = 0$ exceeds that of the other two sublevels. As a result, the incoming Poincaré beam predominantly drives transitions between the ground-state sublevel $M_{g} = 0$ and the excited state $M_{e} = 0$. The corresponding steady-state absorption profile of the Poincaré beam is shown in Fig.~\ref{fig:absorption_linear}. In contrast to the previous case with $\theta_{B} = 0^\circ$, this absorption profile exhibits an asymmetrical pattern with respect to the $y$-axis. This behavior can be understood by examining the local polarization and transition amplitudes. For instance, the excited-state population vanishes in the upper part of the absorption profile, particularly at $b = 200\,\mu m$ and $\phi_{b} = 0^\circ$. From our earlier discussion on the unpolarized atomic target, atoms located at these coordinates experience locally elliptically polarized light. Moreover, atoms in this region are eligible for transitions between all three magnetic sublevels in the ground state and the excited state. However, in the current polarized scenario, the Poincaré beam can only drive the transition between $M_{g} = 0$ and $M_{e} = 0$. From Fig.~\ref{fig:transition_amp}(d), we observe that this particular transition exhibits asymmetry across the beam cross-section, with lower values of $|V_{eg}^{(\mathrm{poin})}|$ in the upper half. Consequently, atoms in the upper part of the beam cross-section interact weakly with the Poincaré beam, leading to an \textit{axial asymmetry} in the absorption profile.

Unlike the case for $\theta_{B} = 0^\circ$, atoms at the center of the beam cross-section interact strongly with the Poincaré beam when $\theta_{B} = 90^\circ$. As discussed earlier, at $\bm{b} = 0$, atoms can undergo transitions from all three ground-state magnetic sublevels to the excited state. However, in the current scenario, the Poincaré beam drives only the $M_{g} = 0 \rightarrow M_{e} = 0$ transition. Furthermore, the pump field maintains a high population in the $M_{g} = 0$ sublevel. Therefore, as the system reaches steady state, the excited-state population attains a maximal and constant value at the beam center, indicating strong interaction of atoms at $\bm{b} = 0$ with the Poincaré probe beam.

\begin{figure}
    \centering
    \includegraphics[width=0.45\textwidth]{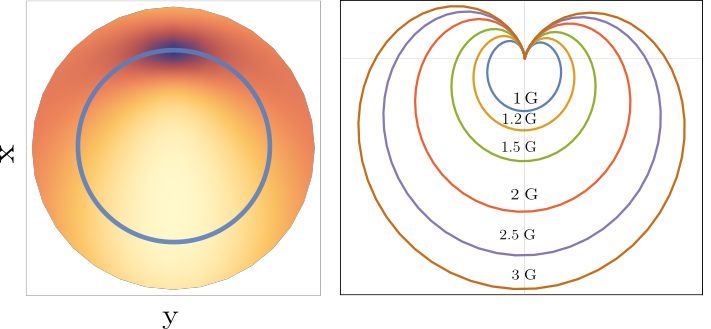}
    \caption{Left: Absorption profile of the Poincaré beam propagating at \(\theta_{B} = 90^\circ\) interacting with a polarized rubidium atomic target immersed in an external constant magnetic field of \(B_{\mathrm{(const)}} = 1\, \mathrm{G}\). Right: Polar plots of the absorption intensity corresponding to the radius \(b = 200\, \mu m\) of the circle shown in the adjacent absorption profile. These polar plots are shown for increasing magnetic field strengths.
}
    \label{fig:sense}
\end{figure}

\subsubsection{Dependence of absorption profile on the strength of the magnetic field}
Next, we examine how the absorption profile of the Poincaré beam interacting with polarized rubidium atoms depends on the strength of the applied magnetic field $B_{\mathrm{(const)}}$. Fig.~\ref{fig:sense} presents the absorption profiles alongside their corresponding polar plots, which exhibit a characteristic \textit{apple-like} pattern. This pattern represents the absorption intensity for atoms located at a fixed radial distance of $b = 200\, \mu m$, marked by a solid circle on the absorption profile. Our analysis focuses on the case $\theta_{B} = 90^\circ$, as this configuration produces an asymmetric absorption profile for the probe light.

The polar plots in Fig.~\ref{fig:sense} illustrate that, as the strength of the magnetic field increases, the \textit{apple-like} pattern expands in size while maintaining its characteristic asymmetry. This behavior suggests a promising method for quantitatively detecting the magnitude of the constant magnetic field by measuring the asymmetry in the absorption profile of the probe light after interaction with the polarized atomic ensemble.

\begin{figure}
    \centering
    \includegraphics[width=0.5\textwidth]{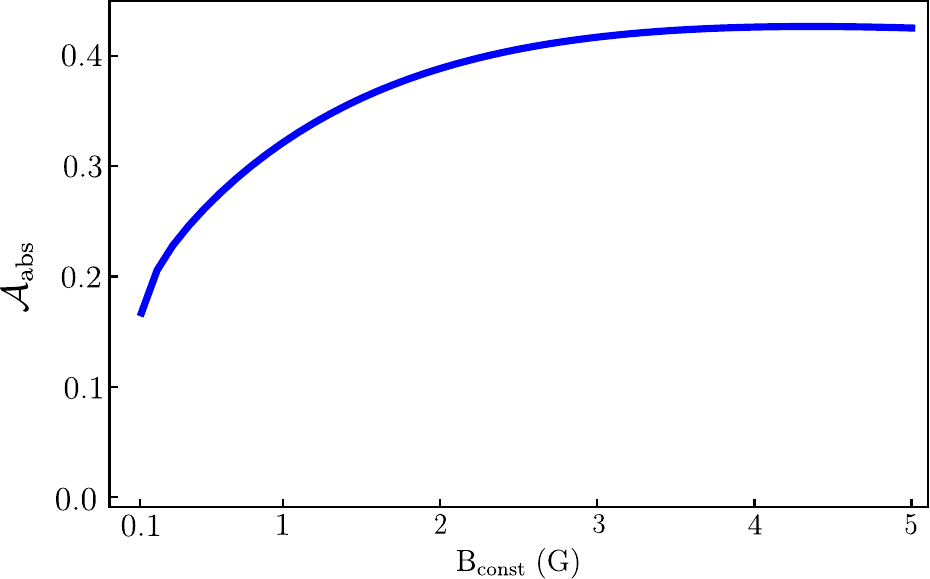}
    \caption{Asymmetry index for the absorption profile of the Poincaré beam propagating at an angle of \(\theta_{B} = 90^\circ\) and interacting with a polarized rubidium atomic ensemble. The asymmetry in the absorption of the light by the atoms is observed to increase with the strength of the applied constant magnetic field in the range of 0.1 to 5 G.}
    \label{fig:asymetry}
\end{figure}

It is well known that any periodic function can be expanded in a Fourier cosine series~\cite{arfken2011/book}. Following this analogy, we expand the density matrix element \(\rho_{ee}\), which depends on the azimuthal coordinate \(\phi_{b}\) at a fixed radial distance \(b = 200\, \mu m\), in terms of cosine functions. This expansion allows us to quantify the asymmetry observed in the excited-state population across the beam cross-section by defining an asymmetry parameter as
\begin{equation}
    \mathcal{A}_{\mathrm{abs}} = \frac{|a_{1}|}{a_{0}},
\end{equation}
where \(a_0\) and \(a_1\) are the zeroth and first Fourier coefficients, respectively. We consider only the first harmonic in the Fourier series because the absorption profile exhibits a pronounced dipolar asymmetry (see Fig.~\ref{fig:absorption_linear}), and the first harmonic coefficient effectively captures this asymmetry. Figure~\ref{fig:asymetry} shows the variation of \(\mathcal{A}_{\mathrm{abs}}\) as a function of the magnetic field strength \(B_{\mathrm{const}}\). The asymmetry parameter ranges between 0 and 1, with 0 indicating perfect symmetry and values closer to 1 indicating stronger asymmetry. For a weak magnetic field of \(0.1\,\mathrm{G}\), \(\mathcal{A}_{\mathrm{abs}} \approx 0.17\), indicating low asymmetry. As the field strength increases, \(\mathcal{A}_{\mathrm{abs}}\) rises, reaching a maximum of approximately 0.42 near \(3.5\, \mathrm{G}\). Beyond this point, the asymmetry parameter saturates, remaining nearly constant.

Physically, this behavior indicates that atoms in the lower half of the beam cross-section interact more strongly with the Poincaré beam and absorb more light compared to those in the upper half. Moreover, the increasing asymmetry with magnetic field strength (also visible in the polar plots of Fig.~\ref{fig:sense}) suggests a direct method to infer the strength of the external magnetic field by analyzing the absorption profile asymmetry of the Poincaré beam interacting with the polarized atomic ensemble.
       
\section{Summary and Outlook}\label{sec:summary}
In this work, we analyzed the interaction between Poincaré beams and an optically polarized atomic target in the presence of constant magnetic field. To conduct this analysis, we used a linearly polarized plane wave as the pump light and Poincaré beam as the probe light. Particularly, we used (paraxial) Bessel modes to construct the vector potential and the corresponding transition amplitude of the Poincaré beam interacting with the target atoms. This interaction process was studied with the help of Liouville-von Neumann equations.

We paid special attention to the population of the excited atomic state, which was subsequently used to plot the absorption profile of the probe Poincaré beam. In particular, we examined this absorption profile both in the absence and presence of the pump light field for \(\theta_{B} = 0^\circ\) and \(90^\circ\), respectively. Our analysis shows that the axial asymmetry present in the polarization profile of the Poincaré beam influences the local transition amplitude, thereby affecting the absorption profiles. Furthermore, we explored the potential application of using the proposed scheme to detect magnetic field strengths by analyzing the asymmetry in the absorption profile of the Poincaré beam.

To simplify our theoretical calculations, we made the following assumptions: (a) thermal motion of atoms in the medium was neglected, (b) the pump and probe light fields were assumed to have same frequency, and (c) one-dimensional constant magnetic field was considered. In the forthcoming publication, we aim to extend this analysis of detecting constant magnetic field by incorporating the effects of atomic thermal motion and a time-dependent magnetic fields.

\section*{Acknowledgments}
We acknowledge support from the Research School of Advanced Photon Science of the Helmholtz Institute Jena.



\bibliography{Bibliography}

\end{document}